\documentclass[a4paper,11pt]{article}

\usepackage{pos}
\usepackage{lipsum} 
\usepackage{subcaption}
\usepackage{lineno}

\title{The Optical Sensor for IceCube-Gen2}

\ShortTitle{Optical Sensor for IceCube-Gen2}

\author{The IceCube-Gen2 Collaboration \\{\normalsize \normalfont(a complete list of authors can be found at the end of the proceedings)}\\}

\emailAdd{alexander.kappes@uni-muenster.de}

\abstract{

An innovative optical module (OM) with segmented light-sensitive area has been developed for IceCube-Gen2 that will take neutrino astronomy at the South Pole to the next level. It builds on the successful features of the mDOM and D-Egg modules of IceCube Upgrade while adapting to the smaller borehole diameter of IceCube-Gen2. The newly developed OM, which is being tested in IceCube Upgrade, serves as a prototype for the planned mass production of about 10,000 OMs for IceCube-Gen2. To simplify the assembly process, important changes were made to the design, in particular to integrate the new gel pad concept. This replaces the 3D-printed support structure of the mDOM while maintaining through total internal reflection the increased light collection efficiency of the reflector rings. In addition, the design features local generation of high voltage for each photomultiplier tube (PMT) via a Cockcroft-Walton circuit and the full digitization of the signal on each PMT base with a sampling rate of 60 MSpS. This significantly reduces the complexity of the mainboard so that it fits into the limited space available. This article describes the development status and presents the performance of the first prototypes.

\vspace{4mm}

{\bfseries Corresponding author:}
A.~Kappes$^{1*}$\\
{$^{1}$ \itshape Institute for Nuclear Physics, University Münster}\\
$^*$ Presenter
}

\ConferenceLogo{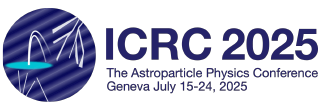}

\FullConference{39th International Cosmic Ray Conference (ICRC2025)\\
 15–24 July 2025\\
Geneva, Switzerland\\}

\begin{document}

\maketitle

\section{Introduction}\label{sec:intro}
The IceCube Neutrino Observatory \cite{Aartsen:2016nxy} has been successfully operating in its full configuration for almost 15 years and is characterized by a remarkably high stability and up-time. This is due in particular to the Digital Optical Module (DOM) \cite{Aartsen:2016nxy}, which has proven to be extremely robust and reliable. Of the 5160 DOMs, some of which have been in the ice for 20 years, only 2\% have ever permanently failed. Since its completion in 2011, IceCube has made many groundbreaking observations, such as the first detection of a high-energy diffuse cosmic neutrino flux \cite{IceCube:2014stg}, the identification of the AGN NGC\,1068 \cite{IceCube:2022der} as a steady source of high-energy neutrino emission, and the observation of neutrinos from the Milky Way \cite{IceCube:2023ame}. To take neutrino astronomy at the South Pole to the next level, a major expansion of the existing detector is being planned which will include a significantly enlarged optical array in the deep ice together with a radio array and other surface extensions \cite{IceCube-Gen2:2020qha}. The optical array of IceCube-Gen2 will consist of around 120 new strings with almost 10,000 newly developed optical sensors, which are intended to achieve an effective light-sensitive area four times larger than the DOMs of the original detector. At the same time, the design must fit into the power budget of $< 4\,\mathrm{W}$ per module and take into account the shrinking diameter of the boreholes which enables faster drilling with lower fuel consumption. And all this while maintaining the high reliability of the original DOMs and keeping within budget.

\section{Sensor design}\label{sec:design}
The sensor development for IceCube-Gen2 builds on the designs of the mDOM \cite{IceCube:2021eij} and D-Egg  \cite{IceCube:2023rfc} for IceCube Upgrade \cite{Ishihara:2019aao} and continues their innovations such as segmented photocathode surface, highly integrated active base and low background, with a particular focus on mass production of nearly 10,000 modules with over 170,000 photomultipliers (PMTs). As reliability in the deep ice, where the modules cannot be repaired, is of paramount importance, the design will be tested in-situ with 12 prototypes to be deployed in IceCube Upgrade in the 2025/26 austral summer season.

\begin{figure}[t!]
\strut\hfill
\includegraphics[width=0.45\linewidth]{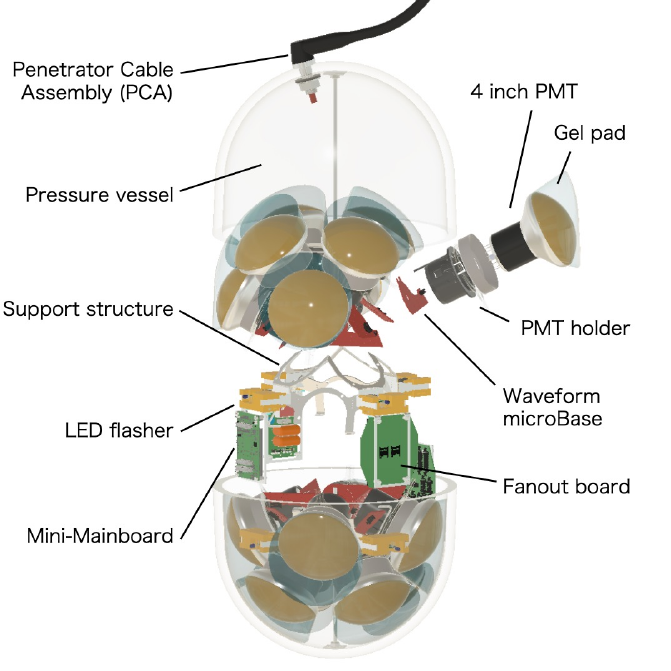}
\hfill
\includegraphics[width=0.22\linewidth]{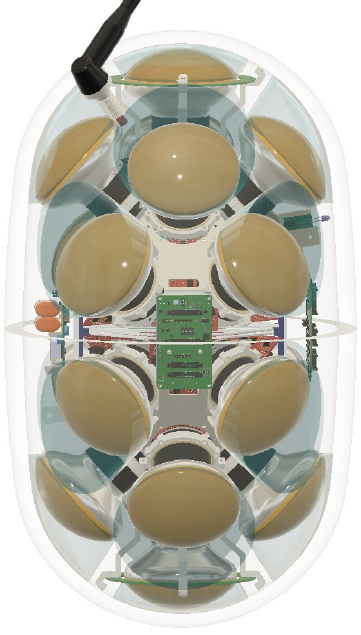}
\hfill\strut
\caption{CAD drawings of Gen2DC-16 (left) and Gen2DC-18 (right).}\label{fig:lom_cads}
\end{figure}

To address the major challenge of achieving an effective area four times larger than that of the IceCube DOM with a smaller module diameter, two different geometric approaches were initially deliberately pursued, which nevertheless have shared the same main components such as photomultiplier, PMT bases and mainboard from the outset. One approach utilizes a pressure vessel designed and produced by Nautilus\footnote{https://nautilus-gmbh.com/en} with an outer diameter of $312\,\mathrm{mm}$ and a height of $444\, \mathrm{mm}$ fitting 16 4"-PMTs. The other has a more elongated shape with diameter $318\,\mathrm{mm}$ and height $540\,\mathrm{mm}$ produced by Okamoto\footnote{https://ogc-jp.com/en/} which can hold 18 4"-PMTs (see Fig.~\ref{fig:lom_cads}). Following the prototyping phase, which includes the deployment of six modules of each design as part of IceCube Upgrade, a final, unique design for the optical module of IceCube-Gen2 is currently being developed and optimized for mass production. The major components of the IceCube-Gen2 DOM design candidate with 16 and 18 PMTs, respectively (hereafter named Gen2DC-16/18), are discussed below, with a focus on the innovative aspects.

\paragraph{Pressure vessel:} The pressure vessel protects the inner components from the high environmental pressure in the deep ice which has been measured to increase to $550\,\mathrm{bar}$ during freeze-in in IceCube-Gen1 deployment. Its diameter is limited by the diameter of the borehole to about $12.5 \,\mathrm{in}$ for IceCube-Gen2. The height is then determined by the pressure rating in combination with the glass thickness. Pressure vessels are made from borosilicate glass which has very good compressive strength and a transmissivity range that matches the PMT wavelength range very well. A limiting factor for the size of the pressure vessel is also its weight as the modules have to be handled during deployment at the South Pole under harsh conditions.

Both pressure vessel designs meet these requirements with the Gen2DC-18 design being somewhat on the heavy side at $17\,\mathrm{kg}$ (Gen2DC-16: $10.7\,\mathrm{kg}$). On the other hand, the Okamoto glass has a very low level of radioactive contamination, which results in a low dark rate (see Fig.~\ref{fig:darkrate}), but at the cost of a higher price and possible limitations on the volume of producible vessels per year.

\paragraph{Photomultiplier:} Studies to maximize the light-sensitive area of the optical module showed that short-necked PMTs with a diameter of 4 inches were the best option. As PMTs with such dimensions were not available at the time, we approached two manufacturers, Hamamatsu Photonics K.K. (HPK) and North Night Vision Technologies (NNVT), who proposed initial designs for their new 4 inch models R16293-01-Y001 and N2041, respectively. These were then improved in several iterations following feedback on measured properties from our side. The final designs were characterized in detail and meet all specified requirements \cite{IceCube-Gen2:2023qpa}. The availability of PMTs from two manufacturers significantly increases the flexibility and procurement capacity per year for the IceCube-Gen2 mass production.

\begin{figure}[t!]
\centering
\begin{subfigure}[][4.7cm][b]{0.3\textwidth}
    \centering
    \includegraphics[width=1\linewidth]{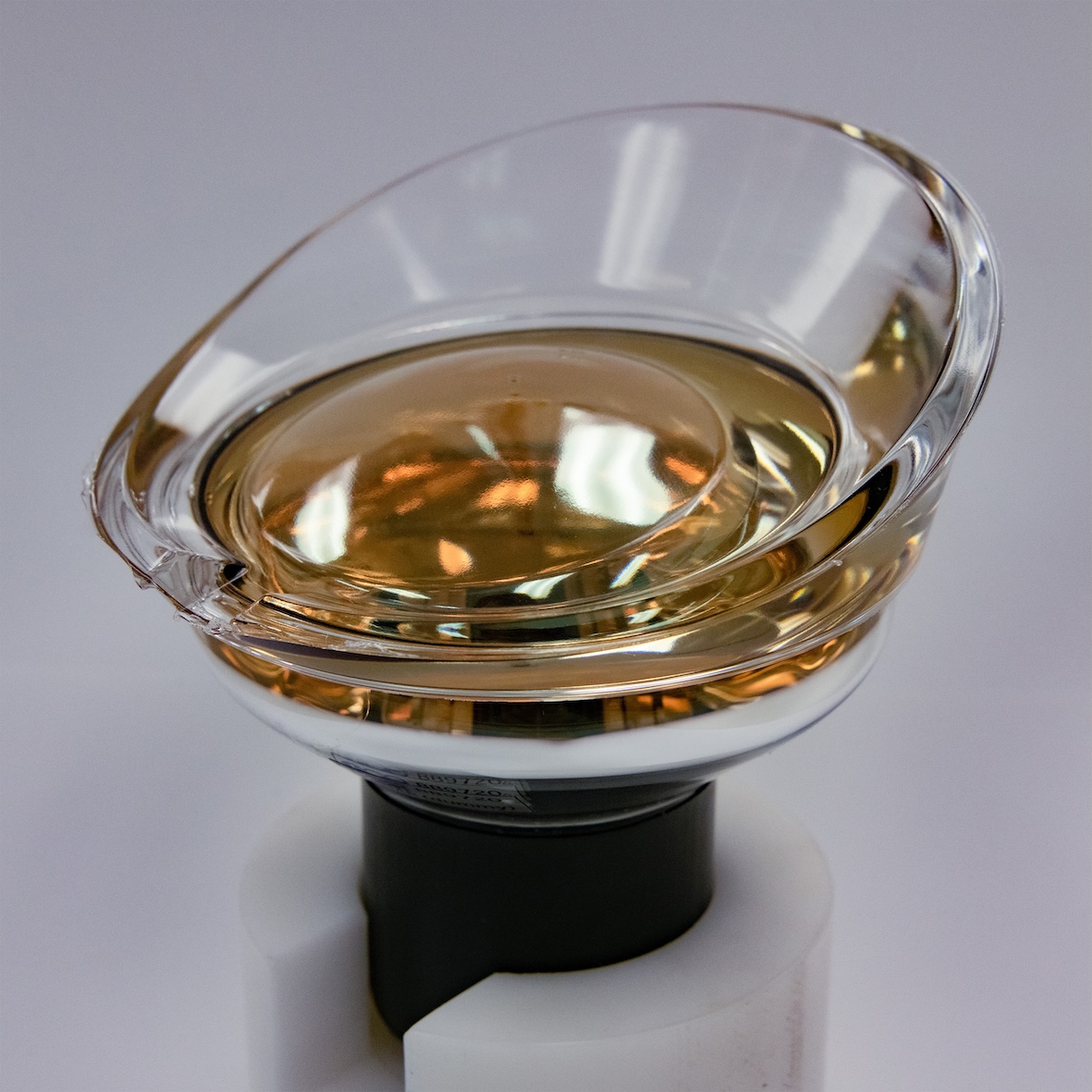}
    \caption{Casted gel pad on a 4 inch PMT.}\label{fig:gelpad}
\end{subfigure}\hfill%
\begin{subfigure}[][7cm][c]{0.65\textwidth}
    \centering
    \includegraphics[width=1\linewidth]{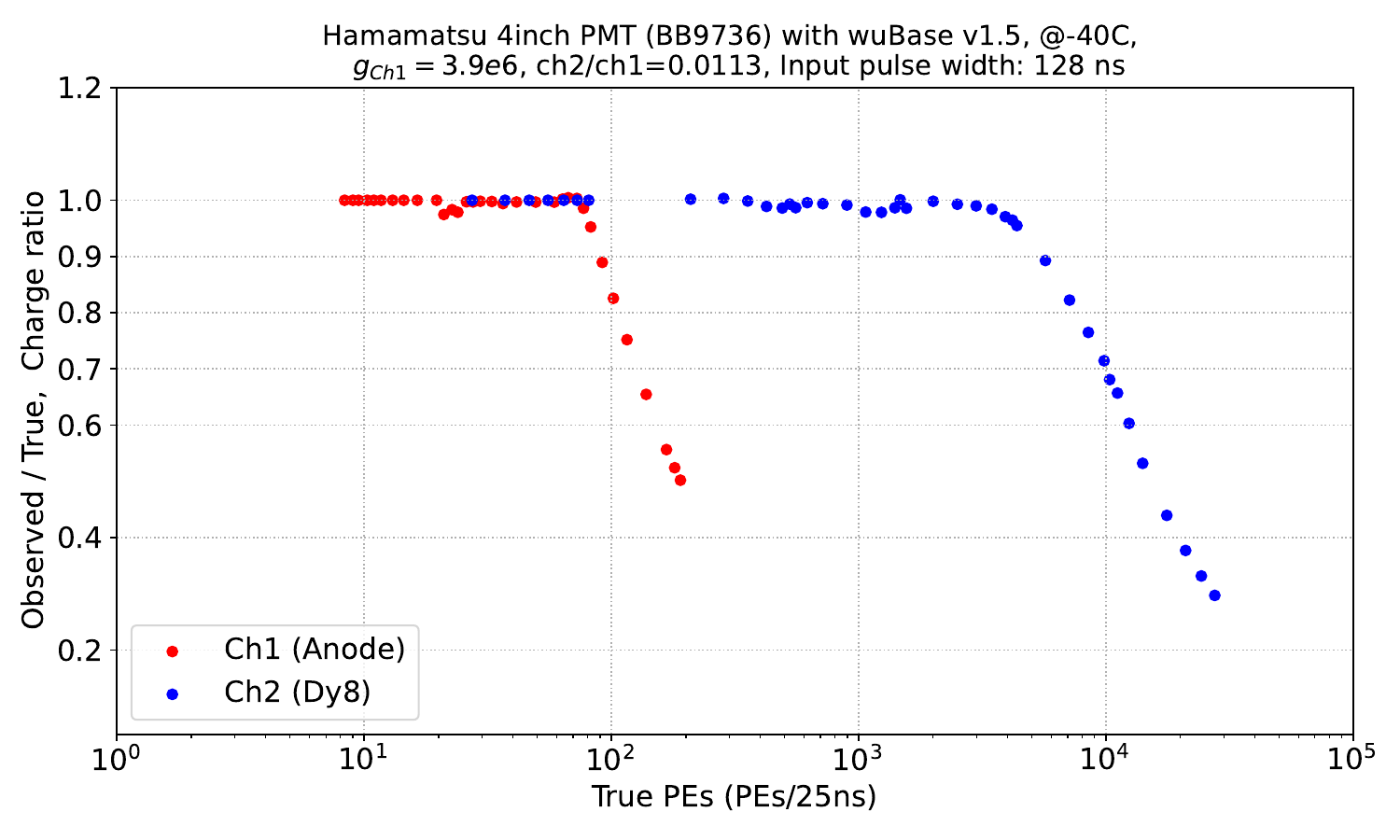}
    \caption{Linear dynamic range of a 4-inch PMT with wuBase. Pulses with charges less than $7\,\mathrm{pe}$ were not measured.}\label{fig:PMT_linearity}
\end{subfigure}
\caption{4" PMT used in the Gen2DC-16 and Gen2DC-18 modules.}
\end{figure}

\paragraph{Optical coupling of PMTs:} In order to minimize photon loss at the transition from pressure vessel to PMTs due to air gaps, the PMTs are coupled to the pressure vessel with a silicone-based optical gel\footnote{Developed for IceCube by Shin Etsu Silicone; http://www.shinetsusilicones.com} which has a very similar refractive index (1.41@$400\,\mathrm{nm}$) as borosilicate glass (1.49@$400\,\mathrm{nm}$). To avoid a hemispherical 3D-printed support structure as in the mDOM and the KM3NeT multi-PMT module, which can induce stress into the surrounding gel layer due to thermal contraction in the ice and, not least, represents a cost factor, we have developed a process in which gel pads are applied to the PMTs before assembly (Fig.~\ref{fig:gelpad}). During module assembly, the PMTs with the pre-casted gel pads are glued to the pressure vessel at the rim with a thin layer of liquid gel while being supported by an inner mechanical structure. The remaining gap between PMT and pressure vessel is later backfilled with gel. We have refined the process to such an extend that the gel pads have an extremely smooth surface that acts like a mirror through total reflection and thus replaces the reflection rings in the mDOM.

\paragraph{PMT bases and mainboard:} Based on the $\mu$Base developed for the mDOM, the concept of moving functionality from the mainboard to the base has been taken a step further. Like the $\mu$Base, the newly developed wuBase features an active Cockcroft-Walton generator for the high voltage based on a $3.3\,\mathrm{V}$ line. In addition, the signal digitizer has now been integrated into the base. The PMT waveform is digitized with a two-channel (low- and high-gain), 12-bit ADC at a rate of $60\,\mathrm{MSps}$. The high-gain channel uses the pulse from the anode whereas the other channel takes the pulse from the 8th (out of 10) dynode. This achieves a high dynamic range of $5,000\,\mathrm{pe}$ (photoelectron) within $25\,\mathrm{ns}$ (Fig.~\ref{fig:PMT_linearity}), which is necessary to avoid saturation during high-energy showers that occur in the vicinity of an optical module. A more detailed description of the PMT base and the electronics of the Gen2DC can be found in \cite{IceCube-Gen2:2023rji}.

Thanks to the outsourcing of these key components to the PMT base, the form factor of the mainboard has been significantly reduced. The so-called Mini-Mainboard measures only $50\times85\,\mathrm{mm^2}$ in size and serves as the command and data processor for a range of IceCube Upgrade devices and also contains a power board. In order to multiplex communication between the Mini-Mainboard and the wuBases, two fan-out boards, one for each hemisphere, have been developed. The fanout boards also provide the electric interface to the LED flasher modules.

\begin{figure}[t!]
\strut\hfill
\includegraphics[height=6.5cm]{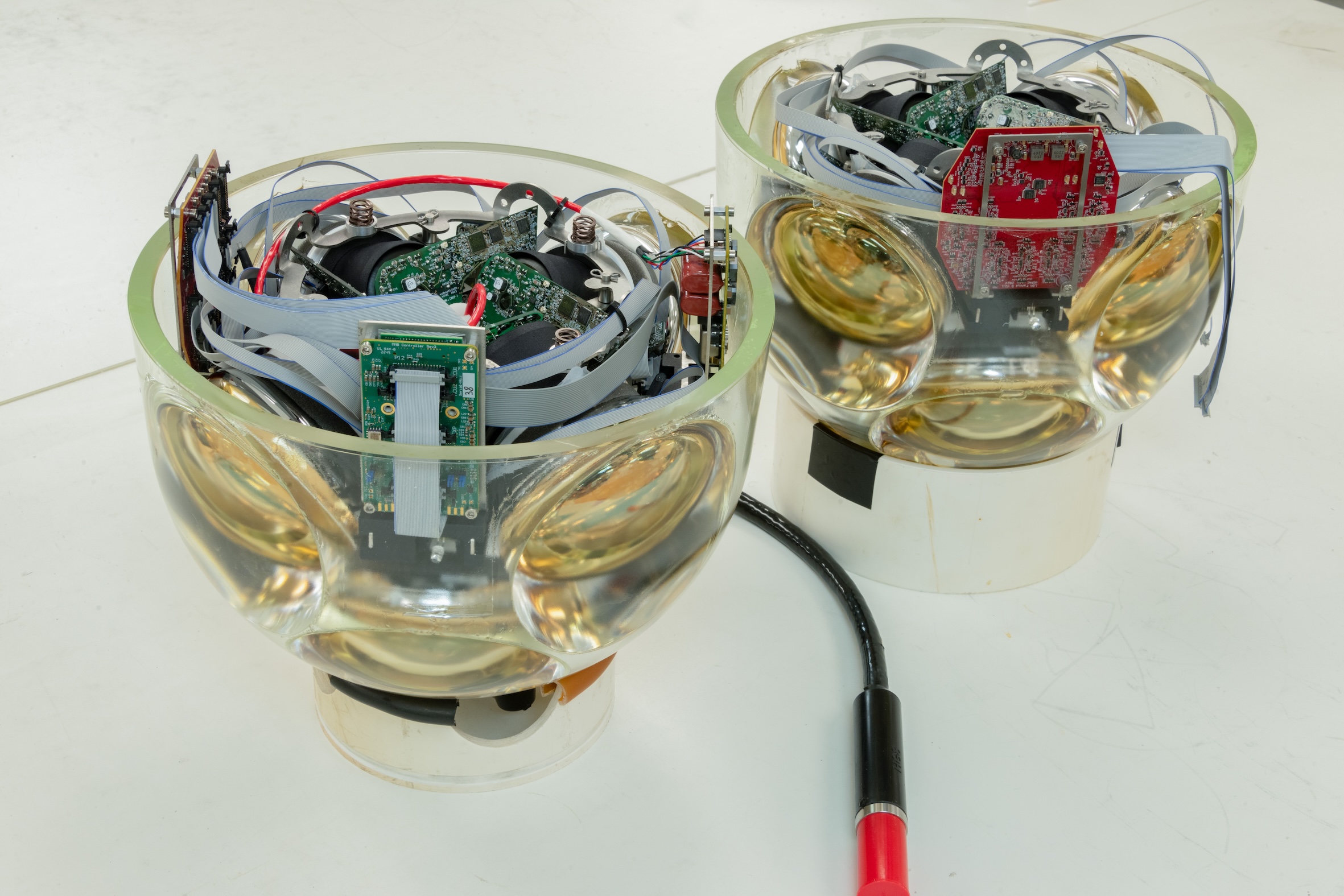}
\hfill
\includegraphics[height=6.5cm]{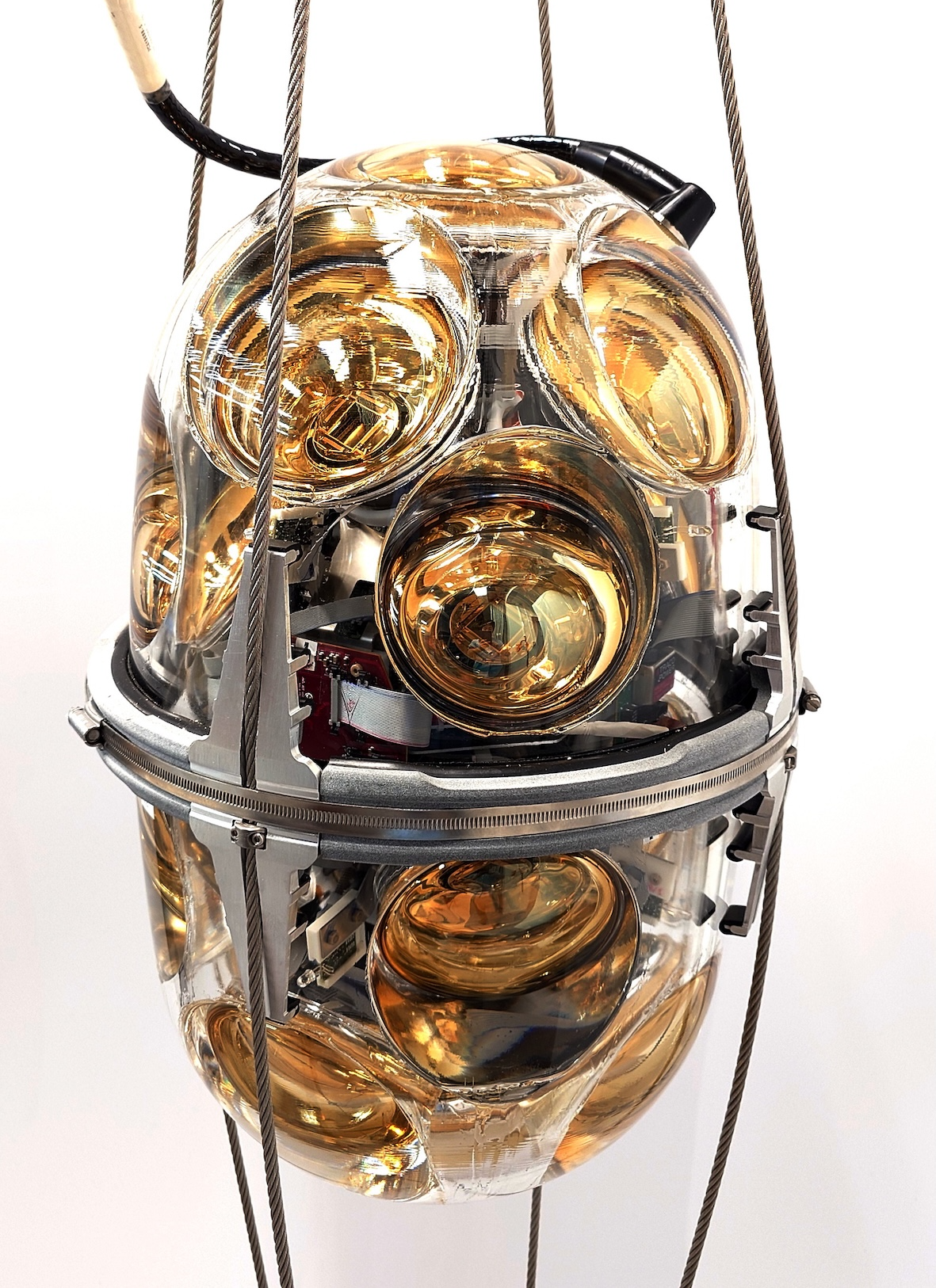}
\hfill\strut
\caption{Left: The two hemispheres of one of the first produced Gen2DC-16 modules with PMTs, electronics and flashers installed. Right: A fully assembled Gen2DC-18 module including harness and penetrator.}\label{fig:lom_prototype}
\end{figure}

\paragraph{Penetrator, harness and calibration devices:} The Gen2DC module is mechanically attached to the main cable via a harness (Fig.~\ref{fig:lom_prototype} right) consisting of a waist band around the equator with attached vertical stainless steel ropes which attach to the next modules or the main cable depending on the installation scenario. Power and communications are provided through a penetrator cable that connects to the main cable. In addition to the PMTs and electronics, the Gen2DC modules also feature eight LED flasher modules which are used for in-ice calibration. 

\section{Current development status}\label{sec:status}
The aim of the parallel development of Gen2DC-16 and Gen2DC-18 is to test different mechanical approaches in order to optimize the photocathode area with regard to the complexity of assembly and the handling of the modules under extreme environmental conditions. A major challenge was the development of a process suitable for mass production for casting high-quality gel pads with smooth surfaces that function effectively as light-collection cones. Fitting the base boards into the tight space was also a major task and motivated the development of the wuBase.

In March 2024, the two Gen2DC designs successfully passed the collaboration internal final design review, and the production of 10 modules of each type was started. Up to now, 9 Gen2DC-16 and 10 Gen2DC-18 have been produced. The production of modules for IceCube Upgrade were finished in Spring 2025. Of each type, 7 modules will be shipped to the South Pole this Summer where 6 will be deployed within IceCube Upgrade in the 2025/26 polar season. The Gen2DC modules will be fully integrated into the IceCube DAQ system which will allow to evaluate their performance using the full IceCube knowledge and resources. 

Parallel to the construction of the modules for IceCube Upgrade, work began on developing the final design for IceCube-Gen2. 

\section{Expected performance and first lab measurements}\label{sec:performance}
\begin{figure}[t!]
\strut\hfill
\includegraphics[height=0.25\textheight]{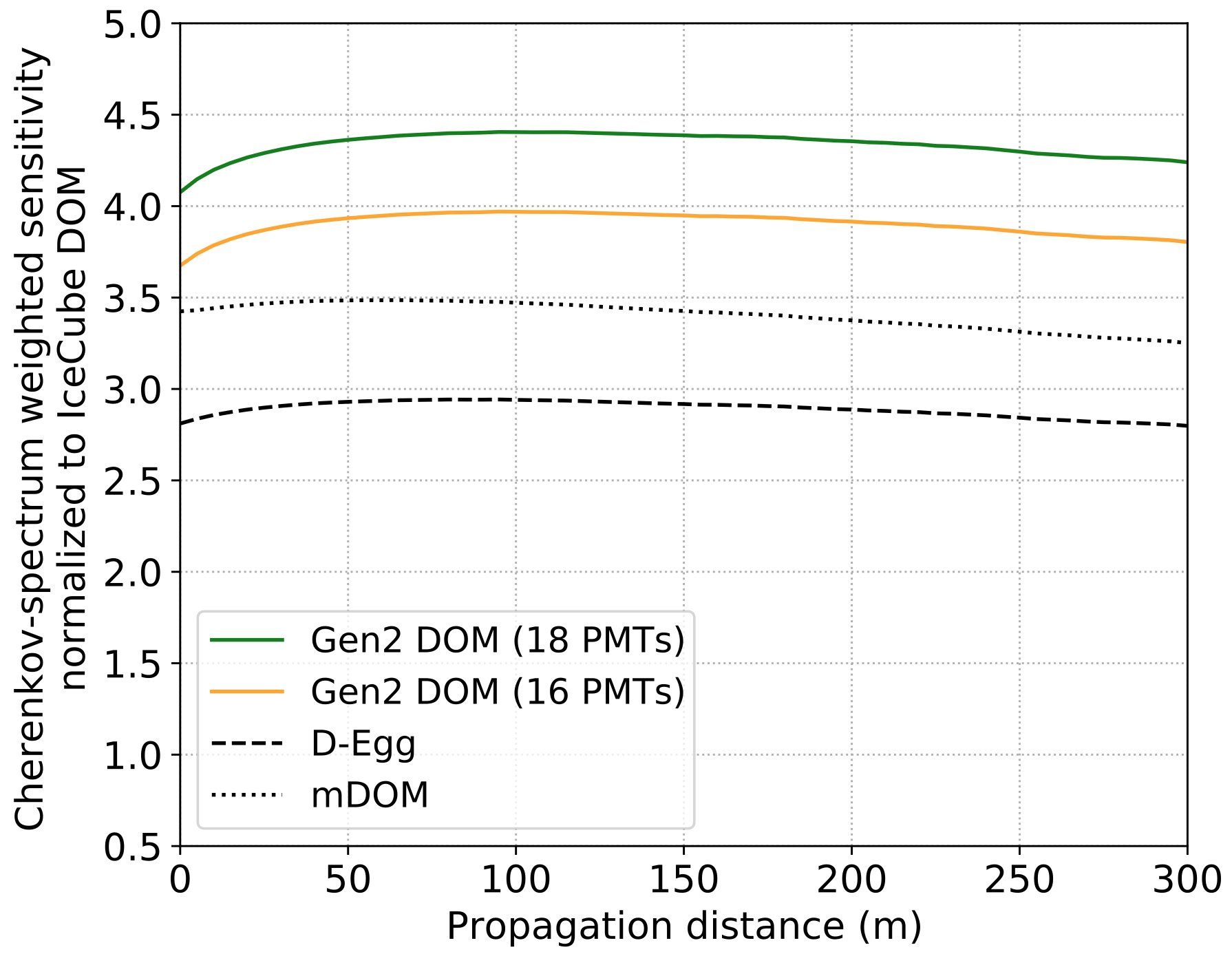}
\hfill
\includegraphics[height=0.25\textheight]{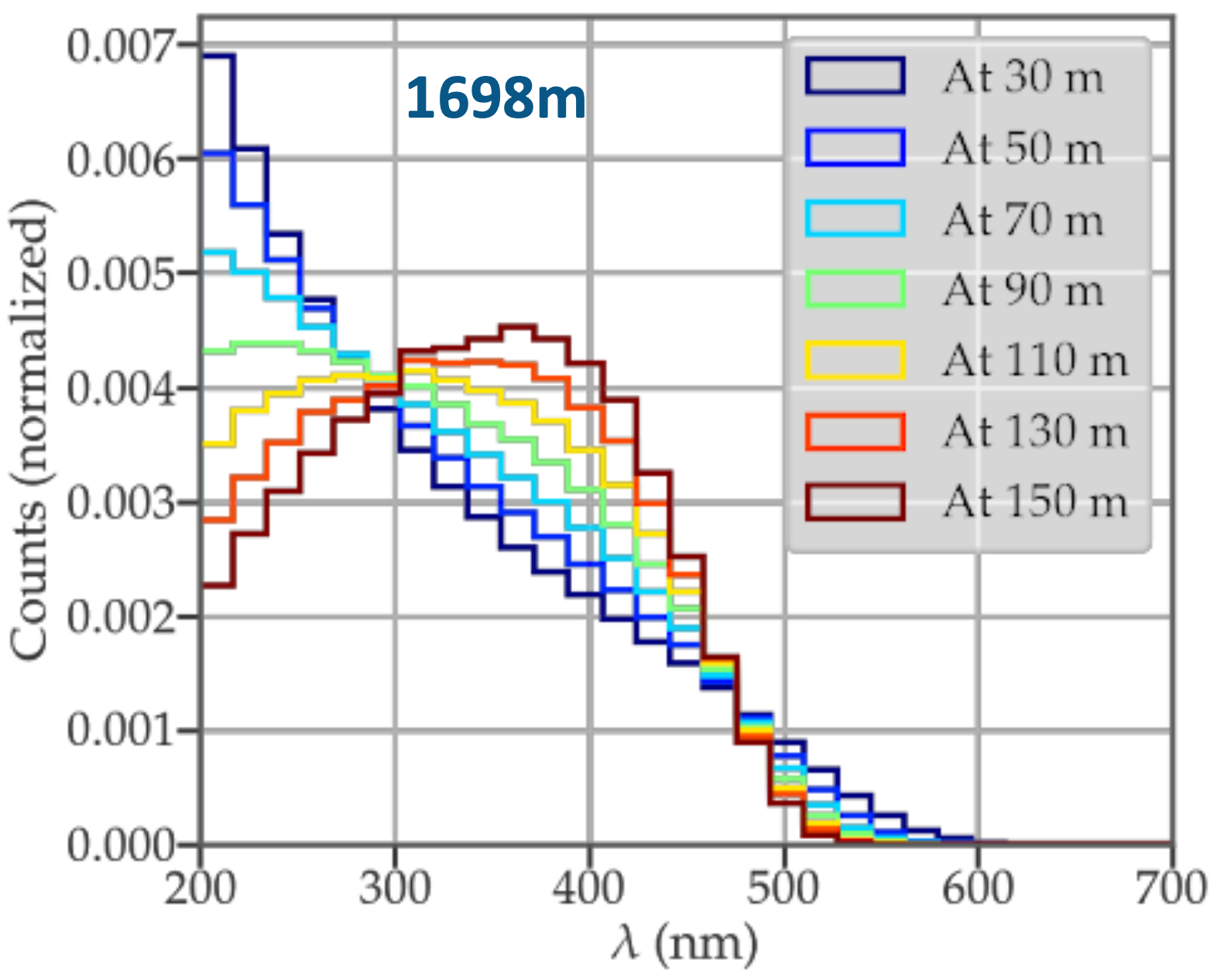}
\hfill\strut
\caption{Left: Simulated module sensitivities of Gen2DC-16 and Gen2DC-18 relative to the current IceCube optical module. Also shown  are the relative sensitivities of mDOM and D-Egg from IceCube Upgrade for comparison. Taken from \cite{Gen2-TDR}. Right: Absorption weighted Cherenkov spectrum in ice at depth $1698\,\mathrm{m}$ as a function of propagation distance.}\label{fig:effarea}
\end{figure}

One of the main goals in the development of the Gen2DC modules was to achieve a light-sensitive area at least four times larger than that of the current IceCube DOM. Figure~\ref{fig:effarea}~left shows the wavelength averaged ratio of the Gen2DC-16 and Gen2DC-18 effective areas relative to that of the current IceCube DOM together with that of other IceCube Upgrade modules using a Geant4 based simulation. As the absorption of the pressure vessel glass and the PMT quantum efficiency strongly depends on the photon wavelength, the effective area needs to be averaged over the detected wavelength spectrum. Though the Cherenkov spectrum continuously rises towards lower wavelength, the wavelength dependent absorption and scattering of photons in the ice leads to a suppression of wavelength below about $300\,\mathrm{nm}$ and above $500\,\mathrm{nm}$ (Fig.~\ref{fig:effarea}~right). Convoluting this propagation dependent spectrum with the effective areas results in the plot on the left side. For IceCube-Gen2 with its probable string distance of about $240\,\mathrm{m}$, both Gen2DC models fulfill the effective area requirement.

\begin{figure}[t!]
\centering
\includegraphics[width=0.48\linewidth]{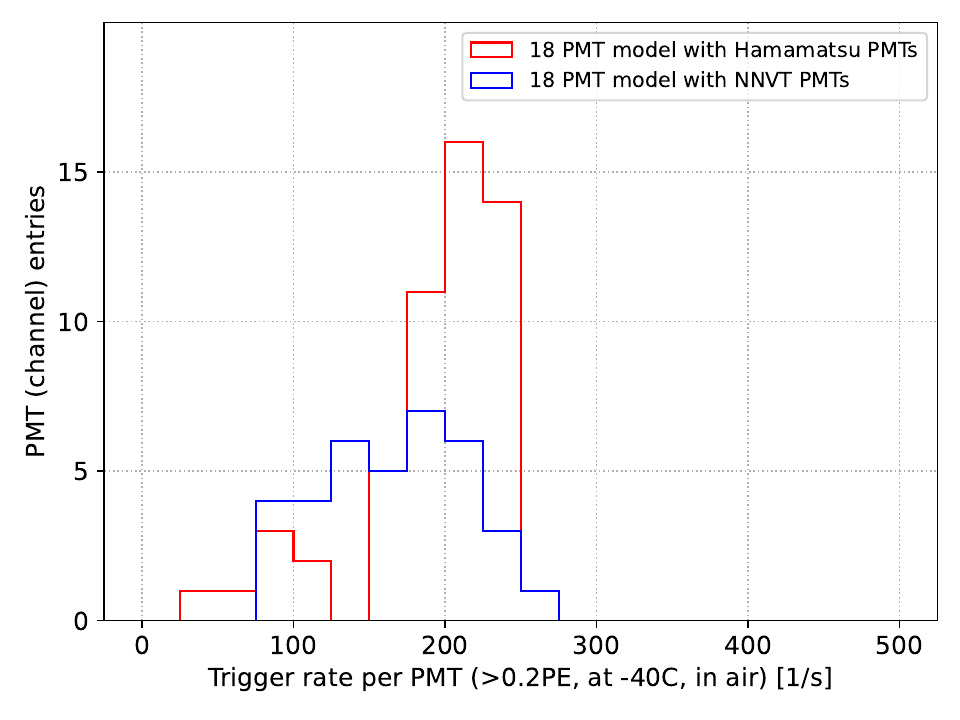}
\caption{Dark rate per PMT (red HPK, blue NNVT) in Gen2DC-18 modules for a threshold of $0.2\,\mathrm{pe}$ at $-40^\circ\mathrm{C}$ in air.}\label{fig:darkrate}
\end{figure}

In first lab measurements with fully integrated Gen2DC modules the dark rates of the PMTs were measured. Figure\,\ref{fig:darkrate} shows the measured dark rates of the PMTs in a Gen2DC-18 module at a threshold of $0.2\,\mathrm{pe}$ at $-40^\circ\,\mathrm{C}$ in air in a dark box. The PMTs from both vendors show similar good performance with dark rates between $100\,\mathrm{Hz}$ and $250\,\mathrm{Hz}$. The rates are dominated by photons generated in decays of radioactive contaminations in the pressure vessel glass. 

\begin{figure}[t!]
\strut\hfill
\includegraphics[width=0.47\linewidth]{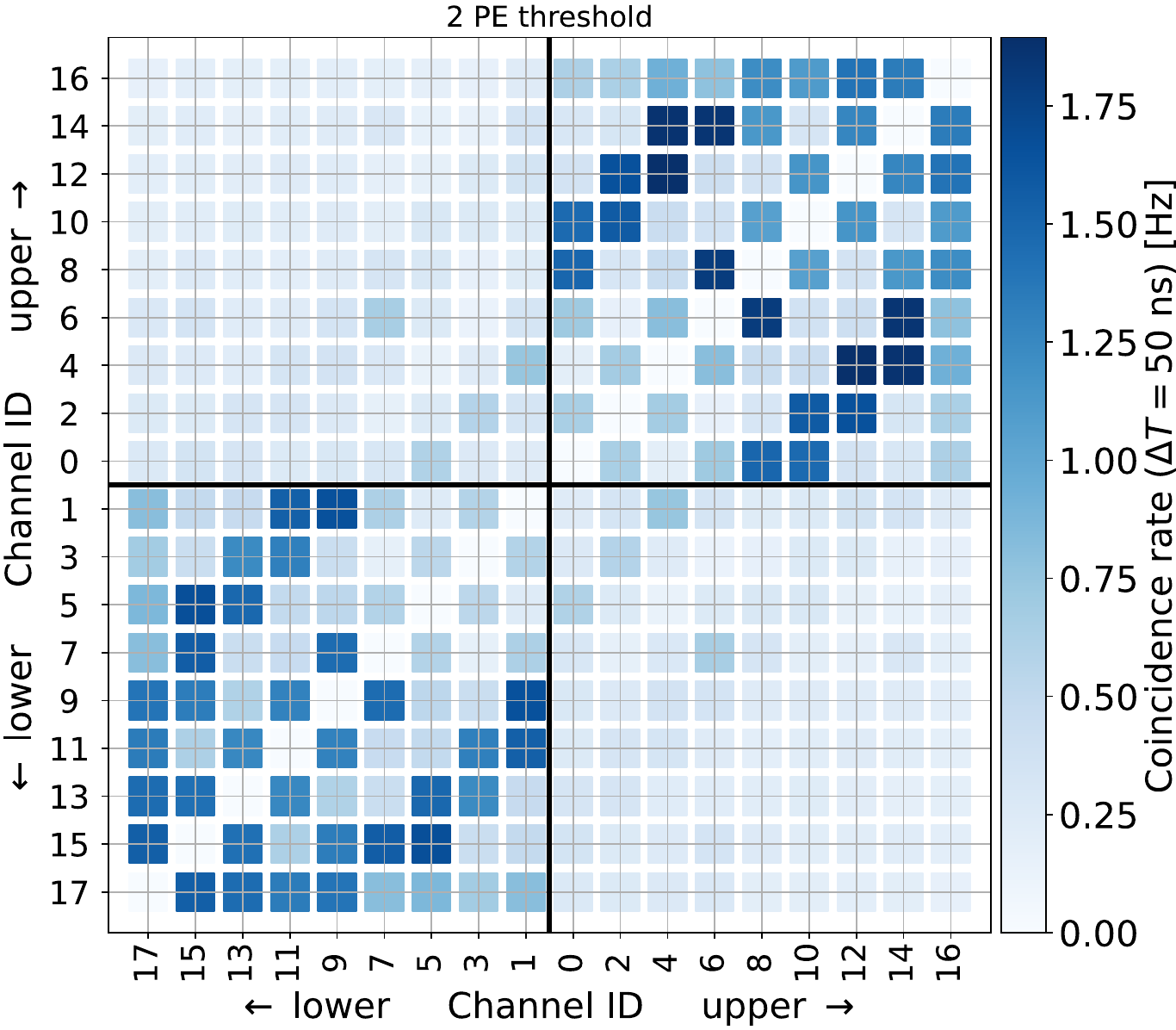}
\hfill
\includegraphics[width=0.5\linewidth]{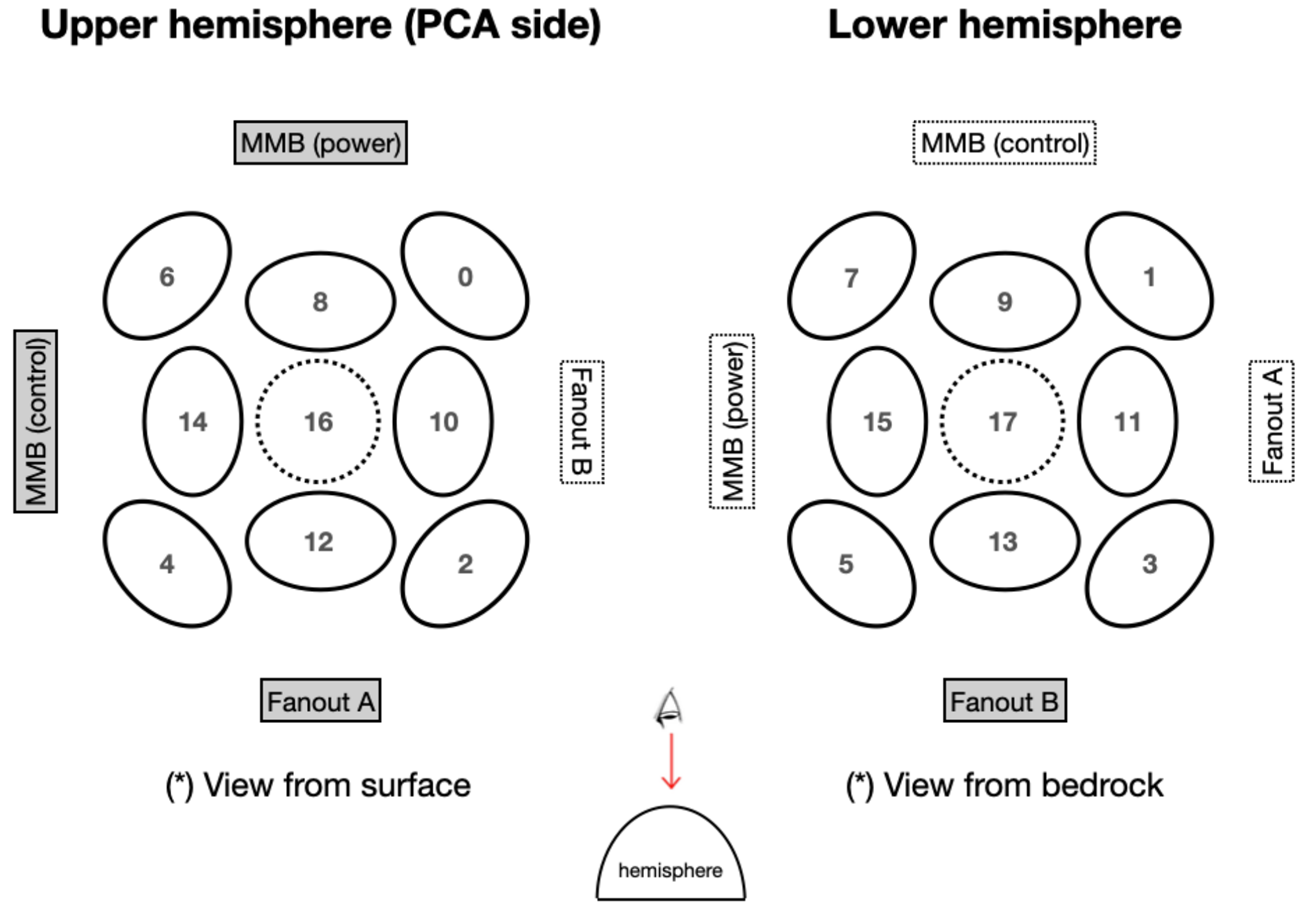}
\hfill\strut
\caption{Left: Coincidence rates in a \emph{taped} Gen2DC-18 between any given pair of PMTs within a time window of $50 \,\mathrm{ns}$ and at a threshold of $2\,\mathrm{pe}$. Right: PMT numbering scheme in the two hemispheres.}\label{fig:corr1}
\end{figure}

In another measurement, the coincidence rates between PMTs in a Gen2DC-18 module in a dark box at a threshold of $2\,\mathrm{pe}$ were investigated (Fig.~\ref{fig:corr1}~left). The module was taped with black tape that absorbs photons traveling outwards in the pressure vessel in order to mimic the effect of a surrounding medium with a similar refractive index as the pressure vessel glass. The location of the PMTs is depicted on the right side of the figure. As expected, neighboring PMTs show an increased coincidence rate and PMTs in different hemispheres generally a very low one. Somewhat higher coincidence rates between upper and lower hemisphere are observed from neighboring equatorial PMTs, e.g.\ PMT\,6 and PMT\,7. 

\begin{figure}[t!]
\strut\hfill
\includegraphics[width=0.42\linewidth]{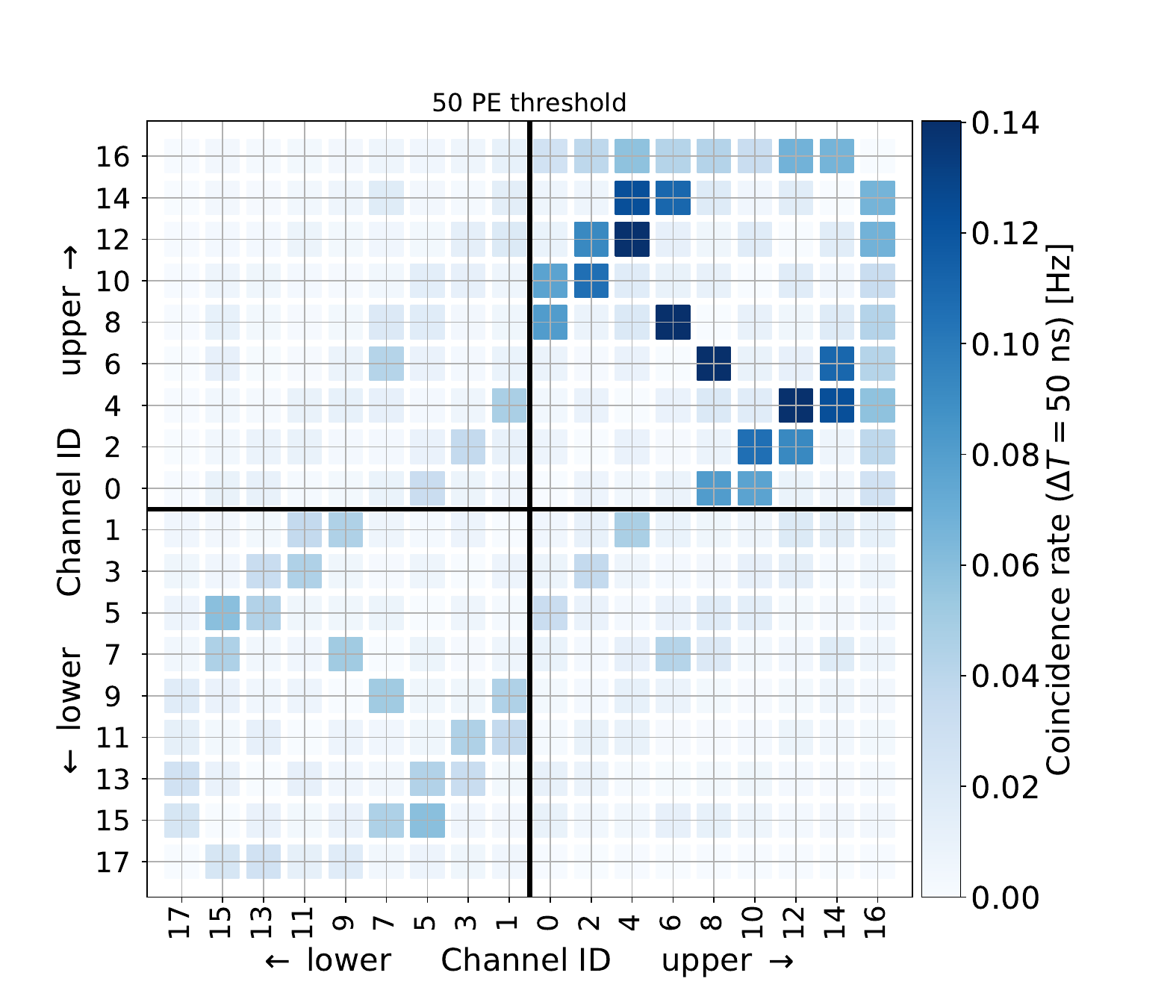}
\hfill
\includegraphics[width=0.55\linewidth]{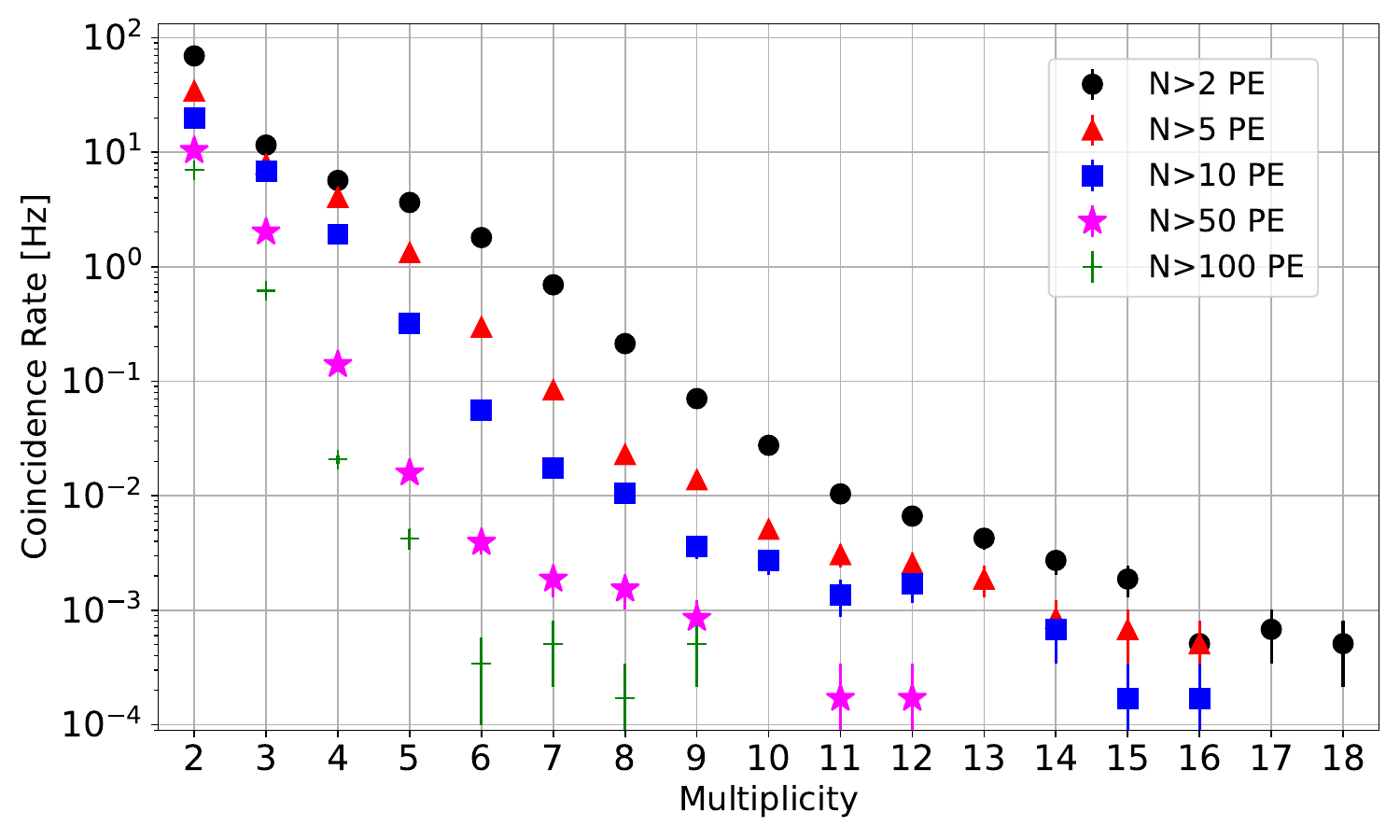}
\hfill\strut
\caption{Left: Coincidence rates in a taped Gen2DC-18 between any given pair of PMTs within a time window of $50 \,\mathrm{ns}$ and at a threshold of $50\,\mathrm{pe}$. For the PMT numbering scheme see Fig.~\ref{fig:corr1}. Right: Coincidence rate as a function of multiplicity for different thresholds for a $50\,\mathrm{ns}$ window .}\label{fig:corr2}
\end{figure}

Next, the threshold was raised from $2\,\mathrm{pe}$ to $50\,\mathrm{pe}$. The resulting coincidence rates are shown in Fig.~\ref{fig:corr2}~left, which now show a clear asymmetry between upper and lower hemisphere for coincidences within the same hemisphere. This is likely caused by atmospheric muons from above: when traversing the pressure vessel glass, they produce a large amount of Cherenkov light in the upper hemisphere that directly reaches the PMTs. On the other hand, when exiting the module on the other side, most of the outward directed muon-induced Cherenkov photons are absorbed by the black tape, which significantly reduces the coincidence rates at high thresholds. Figure~\ref{fig:corr2}~right shows the coincidence rate as a function of multiplicity for thresholds of $2\,\mathrm{pe}$ and $50\,\mathrm{pe}$, respectively. As the threshold increases, high multiplicities are suppressed more than low ones. This supports the above assumption that events with a large number of photons are caused by the Cherenkov light of atmospheric muons, since the photons they produce are deposited only in a small radius around the muon trajectory.

\section{Summary and outlook}\label{sec:summary}
Since the last ICRC in 2023, important progress has been made in the development of a new segmented sensor for IceCube-Gen2. In order to meet the stringent requirements two different designs were initially developed which are based on the gel pad concept for coupling of PMTs to the pressure vessel and contain the same central components (PMTs including base, mainboard, calibration devices). Both designs fulfill all requirements and will be tested in-situ as part of IceCube Upgrade to be installed in the austral summer 2025/26. At the same time, work has started to develop a single design for IceCube-Gen2 where the focus will lie on reliability and manufacturability for the full-scale mass production. 

\bibliographystyle{ICRC}
\setlength{\bibsep}{0pt plus 0.3ex}
\bibliography{references}

%

\clearpage

\section*{Full Author List: IceCube-Gen2 Collaboration}

\scriptsize
\noindent
R. Abbasi$^{16}$,
M. Ackermann$^{76}$,
J. Adams$^{21}$,
S. K. Agarwalla$^{46,\: {\rm a}}$,
J. A. Aguilar$^{10}$,
M. Ahlers$^{25}$,
J.M. Alameddine$^{26}$,
S. Ali$^{39}$,
N. M. Amin$^{52}$,
K. Andeen$^{49}$
% ...
\\
\\
% ...
$^{10}$ Universit{\'e} Libre de Bruxelles, Science Faculty CP230, B-1050 Brussels, Belgium \\
% ...
$^{16}$ Department of Physics, Loyola University Chicago, Chicago, IL 60660, USA \\
% ...
$^{21}$ Dept. of Physics and Astronomy, University of Canterbury, Private Bag 4800, Christchurch, New Zealand \\
% ...
$^{25}$ Niels Bohr Institute, University of Copenhagen, DK-2100 Copenhagen, Denmark \\
$^{26}$ Dept. of Physics, TU Dortmund University, D-44221 Dortmund, Germany \\
% ...
$^{39}$ Dept. of Physics and Astronomy, University of Kansas, Lawrence, KS 66045, USA \\
% ...
$^{46}$ Dept. of Physics and Wisconsin IceCube Particle Astrophysics Center, University of Wisconsin{\textemdash}Madison, Madison, WI 53706, USA \\
% ...
$^{49}$ Department of Physics, Marquette University, Milwaukee, WI 53201, USA \\
% ...
$^{52}$ Bartol Research Institute and Dept. of Physics and Astronomy, University of Delaware, Newark, DE 19716, USA \\
% ...
$^{76}$ Deutsches Elektronen-Synchrotron DESY, Platanenallee 6, D-15738 Zeuthen, Germany \\
$^{\rm a}$ also at Institute of Physics, Sachivalaya Marg, Sainik School Post, Bhubaneswar 751005, India \\
% ...

\subsection*{Acknowledgments}

\noindent
The authors gratefully acknowledge the support from the following agencies and institutions:
USA {\textendash} U.S. National Science Foundation-Office of Polar Programs,
U.S. National Science Foundation-Physics Division,
U.S. National Science Foundation-EPSCoR,
U.S. National Science Foundation-Office of Advanced Cyberinfrastructure,
Wisconsin Alumni Research Foundation,
Center for High Throughput Computing (CHTC) at the University of Wisconsin{\textendash}Madison,
Open Science Grid (OSG),
Partnership to Advance Throughput Computing (PATh),
Advanced Cyberinfrastructure Coordination Ecosystem: Services {\&} Support (ACCESS),
Frontera and Ranch computing project at the Texas Advanced Computing Center,
U.S. Department of Energy-National Energy Research Scientific Computing Center,
Particle astrophysics research computing center at the University of Maryland,
Institute for Cyber-Enabled Research at Michigan State University,
Astroparticle physics computational facility at Marquette University,
NVIDIA Corporation,
and Google Cloud Platform;
Belgium {\textendash} Funds for Scientific Research (FRS-FNRS and FWO),
FWO Odysseus and Big Science programmes,
and Belgian Federal Science Policy Office (Belspo);
Germany {\textendash} Bundesministerium f{\"u}r Forschung, Technologie und Raumfahrt (BMFTR),
Deutsche Forschungsgemeinschaft (DFG),
Helmholtz Alliance for Astroparticle Physics (HAP),
Initiative and Networking Fund of the Helmholtz Association,
Deutsches Elektronen Synchrotron (DESY),
and High Performance Computing cluster of the RWTH Aachen;
Sweden {\textendash} Swedish Research Council,
Swedish Polar Research Secretariat,
Swedish National Infrastructure for Computing (SNIC),
and Knut and Alice Wallenberg Foundation;
European Union {\textendash} EGI Advanced Computing for research;
Australia {\textendash} Australian Research Council;
Canada {\textendash} Natural Sciences and Engineering Research Council of Canada,
Calcul Qu{\'e}bec, Compute Ontario, Canada Foundation for Innovation, WestGrid, and Digital Research Alliance of Canada;
Denmark {\textendash} Villum Fonden, Carlsberg Foundation, and European Commission;
New Zealand {\textendash} Marsden Fund;
Japan {\textendash} Japan Society for Promotion of Science (JSPS)
and Institute for Global Prominent Research (IGPR) of Chiba University;
Korea {\textendash} National Research Foundation of Korea (NRF);
Switzerland {\textendash} Swiss National Science Foundation (SNSF).

\end{document}